\documentclass[preprint,aps,eqsecnum,showpacs,psfig]{revtex4}
\usepackage{graphicx}
\usepackage{latexsym}
\begin{document}
\title{Gravitational lensing in braneworld gravity: formalism and applications}
\author{Supratik Pal \footnote{Electronic address: {\em supratik@iucaa.ernet.in} ;
Present address: Physics and Applied Mathematics Unit, Indian Statistical Institute, 203 B.T.Road, Kolkata 700 108, India}}
${}^{}$
\affiliation{Inter--University Centre for Astronomy and Astrophysics, 
Post Bag 4, Ganeshkhind, Pune 411 007, India}
\author{Sayan Kar \footnote{Electronic address: {\em sayan@cts.iitkgp.ernet.in}}
${}^{}$}
\affiliation{Department of Physics and Centre for
Theoretical Studies, Indian Institute of Technology Kharagpur 721 302, India}
\vspace{.5in}

\begin{abstract}

In this article, we develop a formalism which is different from the 
standard lensing scenario and is necessary for understanding lensing by
gravitational fields which arise as solutions of the effective
Einstein equations on the brane. We obtain general expressions 
for measurable quantities such as time delay, deflection angle, 
Einstein ring and magnification. 
Subsequently, we estimate
the deviations (relative to the standard lensing scenario) 
in the abovementioned quantities  
by considering the line elements for clusters and spiral galaxies 
obtained by solving the effective Einstein equations on the brane. 
Our analysis reveals that 
gravitational lensing can be a useful tool
for testing braneworld gravity as well as the existence of extra dimensions.

\end{abstract}

\pacs{04.50.+h, 95.35.+d, 98.62.Sb}

\maketitle


\section{Introduction}

One of the path-breaking successes of Einstein's general theory of relativity 
is its prediction of  the amount of bending of light by a gravitating object.  
That a light ray can be deflected by the gravitational field of a massive 
object was
indicated, as early as in 1704, by Newton. It was Einstein, however,  who first
used the equivalence principle to calculate this `Newtonian' deflection angle \cite{einlens1}.
Later on, he obtained the formula \cite{einlens2} based on his general relativistic field equations
and found the deflection angle to be exactly twice the Newtonian deflection. This angle,
though very small, was found to be in excellent agreement  in the solar system,
when measured by Eddington and others during a total solar eclipse 
\cite{edding1}.
Eddington, among others, also pointed out the possibility of having multiple
images of a source due to this light bending \cite{edding2}. Later on, Chowlson \cite{chowl}
indicated to the formation of the Einstein ring  by the images for a specific alignment of the source.
This effect was also independently shown by Einstein himself \cite{einring}.
All these properties, resonating with refraction in geometrical optics, led
to the conclusion that {\em a gravitating object can behave like  a lens --
the gravitational lens.}

Because of  excessively small values for the deflection angle, physicists, including Einstein himself,
 were not too sure whether  these properties could be detected some day.
Zwicky, the most enthusiastic among all, calculated the mass of galaxies inside clusters 
by using gravitational lensing 
\cite{zwicky1} and suggested that the magnification of distant fainter galaxies 
can make them visible\cite{zwicky2}. However, physicists had to wait till 1979 for observational
verifications. It was only after   the  discovery of lensing effects by 
the quasar QSO 0957+561A,B \cite{qso} (that they are in fact double images of
a single QSO) when  the predictions of Zwicky and others came out to be  
true. Subsequently, several gravitational lenses have been detected, which have eventually made
the subject an attractive and promising field of research today \cite{sef, narayan, schnei1, safa, kuij}.

Of late, gravitational lensing has emerged as an important probe of structures and has found
several applications in cosmology and astrophysics \cite{schnei2}.
To mention a few, most of the
lens systems involve sources and lenses at moderate or high redshift, 
thereby making it possible to study the geometry of the universe by lensing.
Thus, the Hubble parameter \cite{aphubble}
and the cosmic density can be determined by using multiple-image lens systems and 
time delay between the different light paths of multiply imaged source, such as quasars. 
The quantitative analysis of the multiply imaged
sources  and Einstein radius can provide detailed information 
on the mass of the deflector\cite{qso}, by knowing the angular diameters
and redshifts of the source and the lens.
Further, the magnification
and shear  effects due to weak lensing can be used to obtain statistical 
properties of matter distribution between the observer and the source \cite{sef}. 
So, it can  be used to study the properties of  dark matter halos surrounding
galaxies, and thus, provide a test for its existence. 
The detection of cosmic shear  plays an important role in 
precision cosmology. The arcs, which result from a very strong distortion
of background galaxies,  can be used to constrain cosmological parameters \cite{appara}.
Another interesting application is that
it can serve as a crucial test for any modified theory of gravity.
In \cite{formal1} a rigorous, analytical formalism was developed
in order to study lensing beyond the weak deflection limit--the motivation
there being the search for signatures of modified gravity. This formalism
was further investigated in \cite{formal2} for PPN metrics and then
in \cite{formal3} for metrics that arise in the context of braneworld
gravity. Though not entirely a strong lensing study, the analysis in
\cite{formal1,formal2,formal3} goes much beyond the usual weak deflection
limit. A nice review of the current status of gravitational lensing
beyond the weak field, small angle approximation can be found in
\cite{perlick}.  
   
Lensing characteristics are essentially determined by the 
gravitational potentials. Lensing effects probe the total matter density,
no matter whether it is luminous or dark. 
Gravitational lensing is thus an important tool to test  theories of gravity 
which predict gravitational potentials different from the one in GR. 

In \cite{sbsk} it was shown that in order to consider dark matter with pressure
in galaxy halos, it is necessary to have two gravitational potentials.
In this approach, the weak field equations with the two potentials are first 
solved to obtain the functional forms of the potentials. 
Deflection of light due to such a weakly relativistic (but not Newtonian) 
scenario is then analyzed in the line  elements obtained \cite{sbsk}. 
Subsequent to the work in \cite{sbsk}, in \cite{clust, altdm},
we have demonstrated that bulk--induced extra dimensional 
effects in braneworld gravity can 
provide an alternative to particle
dark matter. It was claimed that one could re--interpret the standard dark 
matter scenario as a purely geometric (necessarily extra dimensional) 
effect rather than due to some invisible material entity. 
Along with the Newtonian potential, this theory
requires the  existence of another potential. These potentials have been
found for spiral galaxies and clusters. One of our
aims in this article is to develop the lensing formalism for 
a weakly relativistic situation where two gravitational
potentials are necessary. This will then be applied to braneworld gravity. 
To illustrate the formalism, we shall estimate some of the observable 
quantities for cluster and galaxy metrics. We will also indicate  possible
links with observational data. It must be mentioned here that there have been
some earlier investigations along somewhat similar lines 
\cite{formal3, skmslens, lensbh, lensdgp, brstrong, harko}. While, in reference \cite{lensbh}, 
the authors study strong lensing by a braneworld black hole, \cite{brstrong} discusses
strong lensing and \cite{harko} analyzes certain aspects for a typical galactic metric in braneworlds.
In \cite{skmslens}, calculations of bending of light in Garriga-Tanaka and tidal charge  metrics
have been done.
\cite{formal3} provides an extensive lensing study with the Garriga-Tanaka metric.  Lensing calculations
in DGP braneworld models are also around \cite{lensdgp}. More recently,
in \cite{shtanov}, the authors have further explored spherically symmetric
line elements (galaxy halos, in particular) in the context of the various
existing effective theories on the brane.  


\section{Bending of light on the brane}


Following \cite{sbsk, clust, altdm}, we express 
a static spherically symmetric metric on the brane in the weak field limit
using isotropic coordinates as
\begin{equation}
d S^2 = - \left(1 + \frac{2\Phi}{c^2}\right) c^2 dt^2 + \left(1 - \frac{2 \Phi - 2\Psi}
{c^2}\right) d \overrightarrow X^2 
\end{equation}
where $\Phi(r)$ is the Newtonian potential and $\Psi(r)$ -- the relativistic
potential -- adds a non-trivial correction to it, characterizing 
braneworld gravity (or, in general situations where pressure terms in
the energy--momentum tensor are important)
and thus, making the theory distinguishable from GR. Note that with 
the intention of studying  optical properties, we have written 
explicitly included the factors of `$c$'in the line element. 

Lensing effects in the above spacetime metric can be expressed in terms of 
an effective refractive index: 
\begin{equation}
 n = 1 + \frac{|2\Phi - \Psi|}{c^2}
\label{n} 
\end{equation}
Thus the refractive index is greater than 1, confirming that a light ray,
analogous to geometrical optics,
passes through the lens slower than the speed of light in vacuum.
Further, this refractive index is related to the corresponding GR value by
\begin{equation}
 n = n_R - \frac{|\Psi|}{c^2}
\end{equation}
Thus the lens on the brane acts as a optically rarer medium than 
a lens in GR.
From now on, we shall assume that the absolute value is implicitly written
whenever we write the potentials.




Since the light speed is reduced inside the lens, there occurs a delay in the 
arrival time of a light signal compared to another signal passing far away
from the lens with a speed $c$. 
This leads to the  time delay of a photon coming from a distant source ($S$), 
propagating through the lens to a distant observer ($O$)  :
\begin{equation} 
 \Delta t = \int_{S}^{O} \frac{2 \Phi- \Psi}{c^3} dl
\label{timedel} 
\end{equation}
where the integral is to be evaluated along the straight line
trajectory between the source and the observer.
Hence a light ray passing through the lens on the brane suffers a time delay
which is less than its GR value, $\Delta t_R$ (the so-called Shapiro time 
delay \cite{sef}) by an amount
\begin{equation}
  \Delta t_R - \Delta t = \frac{1}{c^3} \int_{S}^{O}  |\Psi| dl
\end{equation}
Thus,  an accurate measurement of the time delay can discriminate between the
two theories of gravity, and thus, can test the scenario from observational
ground.
 
The deflection angle,  $\hat \alpha$, of a photon in this gravitational field
 is determined by  the integral of the gradient of the effective refractive
index perpendicular to the light path. This deflection angle can also be
derived by using Fermat's principle, by extremizing the light travel time
from the source to the observer. Thus, we have, 
\begin{equation}
\hat \alpha = - \int_{S}^{O} \hat \nabla_\perp n
= - \int_{S}^{O} \hat \nabla_\perp \left(1 -
\frac{2 \Phi - \Psi}{c^2} \right) dl
\end{equation}
where $\hat{\nabla}_\perp$ denotes 
the derivative in the direction perpendicular to this trajectory. 
Thus, the deflection angle is related to the GR deflection $\hat \alpha_R$   by
\begin{equation}
 \hat \alpha = \hat \alpha_R - \frac{1}{c^2} \int_{S}^{O} 
\hat \nabla_\perp \Psi  dl
= \hat \alpha_R - \hat \alpha_\Psi
\label{def} 
\end{equation}
where the term involving $\Psi$ is the braneworld correction (or a correction in a modified
theory of gravity)  and, for brevity, will be
depicted as $ \hat \alpha_\Psi$ from now on.

What is obvious from the above equation is that a light ray on the brane
is deviated by a smaller amount in comparison with its corresponding GR deflection.
Consequently, it turns out that {\em measuring the deflection angle can serve 
as a crucial test while comparing braneworld gravity effects with those of GR.}


As a useful illustration, let us consider the thin lens scenario. 
  Most of the spherically symmetric objects can be approximated as a thin lens 
for which the Schwarzschild radius is much smaller than the impact parameter,
so that the lens appears to be thin in comparison with the total extent of the
light path. 

The GR deflection of such a lens is given by the `Einstein angle' \cite{sef} 
\begin{equation}
 \hat \alpha_R = \frac{4 G M(\xi)}{c^2 \xi} = \frac{2 R_S}{\xi}
\end{equation}
where $R_S = 2 G M/c^2$ is the Schwarzschild radius of the lens
(for this reason, this type of lens is also called the Schwarzschild lens)
and $M(\xi) = M $ is the constant mass for a point mass source.
Note that the general expression for the mass function is given by
\begin{equation}
M (\xi) = \int \frac{\Sigma (\overrightarrow \xi) (\overrightarrow \xi
- \overrightarrow \xi^{'})}{|\overrightarrow \xi - \overrightarrow \xi^{'}|^2} 
d^2 \overrightarrow \xi^{'}
\end{equation}
in terms of a two-dimensional vector $\overrightarrow \xi$ on the lens plane,
which is basically the distance from the lens center $\xi^{'} = 0$.
This general expression reduces to a constant mass $M(\xi) = M = constant$
for a point mass source.
Hence a thin lens in  braneworld gravity deviates a light ray
by an amount 
\begin{equation}
 \hat \alpha = \frac{4 G M}{c^2 \xi} - \hat \alpha_\Psi 
\label{thindef} 
\end{equation}
which can be subject to observational verification.


\section{Lensing geometry on the brane}


Apart from the time delay and the deflection angle, the other observable properties of a gravitational lens are 
the position of the image 
and the magnification involving convergence and shear. 
In order to find out these quantities, it is customary to obtain the lensing
geometry in terms of the lens equation. Below is a schematic diagram that shows how a 
gravitational lens functions. A light ray, emerging from the source S, is deflected
by an angle $\hat \alpha$ by the lens L and reaches the observer O, resulting in 
the image  at I. The angular positions of the source and the image with respect to the
optical axis of the lens are $\beta$ and $\theta$ respectively. Here $D_{ds}$, $D_d$ and
$D_s$ are the angular diameter distances between source and lens, lens and observer,
and source and observer respectively. 

\begin{figure}[htb]
{\centerline{\includegraphics {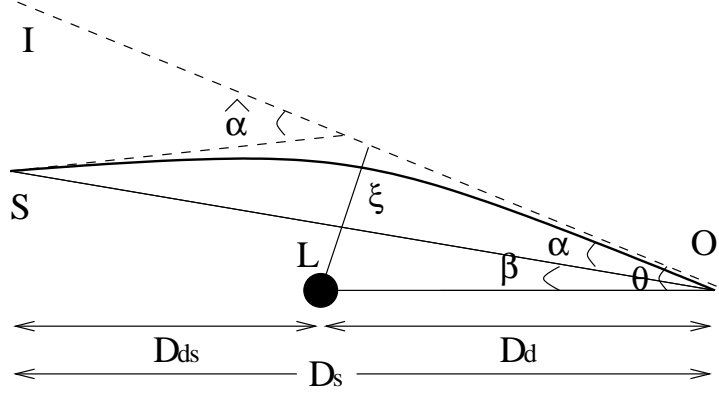}}}
\caption{Schematic diagram of a gravitational lens system}
\end{figure}

Now, the deflection angle being small, the angular positions
bear a simple relation among them. The general lens equation \cite{sef}
reduces to the following:
\begin{equation}
D_s \beta  = D_s \theta - D_{ds} \hat \alpha
\label{lenseq2} 
\end{equation}
Thus, in terms of the reduced deflection angle (where $D_d D_{ds}/ D_s = D$ measures
the effective distance)
\begin{equation}
\alpha = \frac{D_{ds}}{D_s} \hat \alpha = \alpha_R - \alpha_\Psi
\label{reddef} 
\end{equation}
the vector  expression for Eq (\ref{lenseq2}) on the lens plane can be written  as
\begin{equation}
 \overrightarrow \beta = \overrightarrow \theta - \overrightarrow \alpha(\theta)
\label{lenseq} 
\end{equation}
This is  the desired lens equation -- also called the `ray-tracing equation'.
Note that though this equation resembles the lens 
equation in GR, quantitatively this is a different equation, since the 
deflection
angle and the  angular positions in the braneworld gravity are different from
their GR values. This will be revealed from the new results obtained in the subsequent sections.


\subsection{Image formation and Einstein ring}

Equipped with the lens equation, one can now study the formation of images, which will
eventually reveal some interesting facts. 
A source lying on the optical axis ($\beta = 0$) of a circularly symmetric lens
is imaged as a ring, called the Einstein ring \cite{chowl} and the corresponding radius of the ring
is called the Einstein radius. 

The above lens equation (\ref{lenseq}) with two potentials  suggests that the deflection angle $\alpha$
has a modification $\alpha_\Psi$ which is a function of  $\theta$. Hence, 
one needs to know  the exact form of $\Psi$ in 
order to get the correction for a circularly symmetric lens. 
Of course, for the case $\Psi =0$ the results are identical to the GR results,
but this is not so when $\Psi \neq 0$. Below we shall illustrate the
situation with a specific example.

Let us consider the lensing scenario for the Garriga-Tanaka metric which
incorporates the effects of extra dimensions in the exterior gravitational
field of a spherically symmetric, static massive object living on 
the brane \cite{gt}. 
The light bending angle in this metric has been calculated in \cite{skmslens}. 
It is
a straightforward exercise to show that this  metric can indeed be cast into
the form with two potentials  $\Phi$ and $\Psi$ as being discussed in the
present paper. With this new formalism, the net deflection is the
same as obtained in \cite{skmslens}. Explicitly,
\begin{equation}
\hat \alpha = \frac{4GM}{c^2 r} + \frac{4GM l^2}{c^2 r^3}
\end{equation}
where the last term in RHS is the braneworld modification (or, more generally,
a modification due to a second potential).
For this deflection, we can now rewrite the lens equation (\ref{lenseq})
in the form
\begin{equation}
\theta^4 - \beta \theta^3 - \theta_{ER} \theta^2 - (\theta_l)^2 =0
\label{quartic} 
\end{equation}
where $\theta_l$ is the modification due to the characteristic length scale $l$
of the angular position of the image with
respect to the optical axis of the lens.

To obtain the Einstein ring, we put the condition $\beta =0$ in the lens equation. This results
in the following expression for the image position: 
\begin{equation}
\theta^2 =  \frac{1}{2}[(\theta_{ER} \pm 
\sqrt{(\theta_{ER})^2 + 4 \theta_l^2}]
\label{theta2} 
\end{equation}
 The minus sign is
ruled out because it will give imaginary $\theta$.
Consequently, with the valid solution with the positive sign, we arrive at
the following interesting conclusion: In a theory of gravity
with two potentials, the Einstein ring is indeed formed but
the radius of the Einstein ring is different from the GR radius. 
In order to get the full image structure
one needs to look at the roots of the quartic equation (\ref{quartic}), which is not
a very trivial exercise. Of course, one can solve the quartic equation
and find out the roots depicting the image position for this specific metric
and the solutions will definitely give some new results as obvious from Eq (\ref{theta2})
but the results do not always turn out to be tractable. A second independent approach is
the perturbative analysis following \cite{formal3}. 
However, since the results will vary with the expressions for relativistic 
potential for different
metrics, it is
sufficient to realize that the Einstein ring and image position with 
two potentials will
be different from GR results in general and perform the analysis
afresh with the specific potentials under consideration.
The situation is applicable to models of dark matter with relativistic 
stresses, such as \cite{sbsk}, as well. Thus, our formalism is quite general 
irrespective of whether we are studying braneworlds or not.

However, even without the abovementioned analysis,
it is easy to show that the radius of the Einstein ring will be
larger if we have some conditions on possible additional terms in the
deflection angle. 
Let us assume that with the additional terms arising out of a modified 
deflection
angle, the condition for Einstein ring ($\beta=0$) is of the form:
\begin{equation}
\theta= \frac{\theta_{ER}^2}{\theta} + \theta_{ER}^2 \sum_1^m \frac{a_{(2n+1)}}{\theta^{(2n+1)}}
\end{equation}
where the additional terms are encoded in the second term on the RHS, with
arbitrary coefficients $a_{(2n+1)}$. Keeping only the odd order terms in the 
summation
to make sure that $\beta \rightarrow  -\beta$ implies $\theta \rightarrow -\theta$,
one can rearrange the terms of the above equation to give
\begin{equation}
\frac{\theta}{\theta_{ER}^2} - 1 = \sum_1^m \frac{a_{(2n+1)}}{\theta^{2n}}
\end{equation}
Obviously, the RHS is positive as long as all the coefficients $a_{(2n+1)}$ are positive. 
Consequently, wherever such corrections in the deflection angle arise,
the Einstein radius will be greater than its value obtained without them. 

Thus, following the above analysis, for the Garriga-Tanaka metric, the 
Einstein ring will be larger than the
GR case. This is, in general, true for any such metric
with an additional correction term arising due to
pressure-like effects in the source. No matter whether it arises from 
relativistic stresses
or from braneworld modifications, we will have a similar
conclusion as long as the correction varies as inverse powers of $\theta$.
This is, indeed, an interesting fact from observational point of view
and is a clear distinction between the two theories.

However, it is worthwhile to note from Eq (\ref{theta2}) that, with the present example, 
 a circularly symmetric lens forms two images of the source, lying
on either side. While one image ($\theta_{-}$) lies inside the 
Einstein ring, the other one  ($\theta_{+}$) outside. This is how multiple
images are formed by a gravitational lens. This situation is identical 
to GR.

\subsection{Singular isothermal sphere}

Let us now  discuss the 
image formation by a galaxy modeled as an isothermal sphere. 
The matter constituents of a
galaxy  are considered to be in thermal equilibrium, confined by the
spherically symmetric gravitational potential of the galaxy, which 
behaves like a singular isothermal sphere obeying the equation
\begin{equation}
m \sigma_v^2 = k T
\end{equation}
where $\sigma_v$ is the line-of-sight velocity dispersion of the stars and HI clouds 
rotating inside the galaxy. By utilizing the properties of hydrostatic equilibrium
and the velocity profile of HI clouds inside galaxies,
one can easily derive the relation
\begin{equation}
 v_c^2 (r) = \frac{G M(r)}{r} = 2 \sigma_v^2 
\label{veldisperion} 
\end{equation}
which reproduces the observed flat rotation curve. Consequently, under the thin lens
approximation, Eq (\ref{thindef}) implies that  a light ray on the brane
is deflected by an isothermal spherical galaxy by an angle
\begin{equation}
 \hat \alpha = \frac{4 \pi \sigma_v^2}{c^2} - \hat \alpha_\Psi
\label{isodef} 
\end{equation}
Thus, for $\Psi \neq 0$, there is a non-trivial  modification
that tends to alter the GR results. 
Once again the results will differ from GR due to the presence of
a nonzero $\alpha_{\Psi}$ in the above equation. However, as discussed earlier,
the quantitative results will depend exclusively on the specific expression for 
the relativistic potential $\Psi$.


\section{Magnification in braneworld gravity}

 As in geometrical optics, a source not only gets multiply imaged 
by a gravitational lens but the deflected light rays can also change the 
shape and size of
the image compared to the actual shape and size of the source. This happens due to the
distortion of the cross-section of light bundles that changes the solid angle viewed
from the location of the observer. However, the surface brightness of the source
is not affected by the lens as light neither gets absorbed nor emitted during deflection
by the lens.

The quantity representing this change in shape and size of the image with respect to the source 
is called the \textit{magnification} which is given as: 
\begin{equation}
 \mu = \text{det} {\cal M} = \frac{1}{\text{det} {\cal A}}
 \label{invjac} 
\end{equation}
where ${\cal A}$ is the Jacobian of the lens--mapping matrix. 
Below we discuss in detail how to describe and estimate the magnification for
metrics in braneworld gravity.

\subsection{Lensing potential}

\noindent The Jacobian matrix can be expressed conveniently in terms of a 
scalar potential, called the lensing potential, which provides useful physical
insight. 
With a non-zero relativistic potential, the lensing potential is now modified 
to
\begin{equation}
V (\theta) = \frac{D_{ds}}{D_d D_s} \int \frac{2\Phi - \Psi}{c^2} ~dl
\end{equation}
For $\Psi=0$ we get back the Newtonian potential.
Hence, in braneworld gravity, the lensing potential is now reduced by an amount
\begin{equation}
V_\Psi = \frac{D_{ds}}{D_d D_s} \int \frac{\Psi}{c^2} ~dl
\label{vpsi} 
\end{equation}
It is worthwhile to mention two important properties of the lensing potential :
\begin{itemize}
\item[(i)] The gradient of $V$ w.r.t. $\theta$ is the reduced deflection angle 
on the brane
\begin{equation}
  \nabla_\theta V = \frac{D_{ds}}{D_s} \int \hat\nabla_\perp
\left(\frac{2\Phi - \Psi}{c^2} \right) ~dl 
= \alpha
\label{grad} 
\end{equation}
which, together with the GR result $ \nabla_\theta V_R = \alpha_R$, implies
\begin{equation}
  \nabla_\theta V_\Psi = \alpha_\Psi 
\end{equation}

\item[(ii)] The Laplacian of $V$ w.r.t. $\theta$ is the scaled surface mass density
\begin{equation}
  \nabla_\theta^2 V = \frac{D_{ds}}{D_s} \int \nabla_\perp^2
\left(\frac{2\Phi - \Psi}{c^2} \right) ~dl 
= 2 \frac{\Sigma(\theta)}{\Sigma_{\text{cr}}}
\label{lap} 
\end{equation}
where $\Sigma$ is the surface density as already defined and $\Sigma_{\text{cr}}
= (c^2/4 \pi G)(D_s/ D_d D_{ds})$ is its critical value. The scaled surface density,
called the convergence $\kappa$, reveals that $V$ satisfies 2D Poisson equation
\begin{equation}
 \nabla_\theta^2 V = 2 \kappa
\end{equation}
\end{itemize}
It is straightforward to verify that equations (\ref{grad}) and (\ref{lap}) together 
gives the same deflection angle as calculated for a thin lens.


\subsection{Convergence and shear}

\noindent Using the lensing potential, the Jacobian matrix can be written as
\begin{equation}
{\cal A} = \delta_{ij} - \frac{\partial^2(V_R - V_\Psi)}{\partial \theta_i \partial \theta_j}
\end{equation}
wherefrom the inverse of  the magnification tensor turns out to be
\begin{equation}
{\cal M}^{-1} = {\cal M}_R^{-1} + \frac{\partial^2 V_\Psi}{\partial \theta_i \partial \theta_j}
\end{equation}
and the total magnification is given by 
\begin{equation}
\mu = \text{det}{\cal M} = \mu_R \left[1 +\mu_R ~\text{det}
\left(\frac{\partial^2 V_\Psi}{\partial \theta_i \partial \theta_j}\right)\right]^{-1} 
\label{mag2} 
\end{equation}
where $\mu_R$ is the magnification calculated from GR. Clearly, the
magnification in braneworld gravity is different from the corresponding GR value
due to the presence of the additional term inside the square bracket. 
However, in order to comment conclusively on whether the magnification will be
more or less than the GR value, one needs to have a specific expression for
$\Psi$ and check whether the determinant of  the potential due to that  $\Psi$
has a positive or a negative contribution.
In what follows we shall illustrate this situation in a bit more detail.
From now on, we shall use 
$\partial^2 V/ \partial \theta_i \partial \theta_j = V_{ij}$ for brevity.

Two important quantities derived from the linear combinations of the 
components of the Jacobian matrix 
provide the real picture of how a source is mapped onto the image. They are :
\begin{itemize}
\item[(i)] Convergence  $\kappa = \frac{1}{2} (V_{11} + V_{22})
= \frac{1}{2} \text{Tr} V_{ij}$ 
\item[(ii)] Shear $\gamma = \sqrt{\gamma_{1}^{2} +\gamma_{2}^{2}}$ 
where $\gamma_{1} = \frac{1}{2} (V_{11} - V_{22}) = \gamma \cos 2 \phi$ 
 and $\gamma_{2} = V_{12} = V_{21} = \gamma \sin 2 \phi$ 
\end{itemize}
The first one depicts the change in size of the source 
when imaged while the latter one gives the change in shape. A combination of the
two accounts for the total magnification.
In terms of convergence and shear, the Jacobian matrix can be expressed as 
\begin{equation}
 {\cal A} = 
\left( \begin{array} {cc}
  1 - \kappa - \gamma_{1}& -\gamma_{2} \\
  - \gamma_{2}&  1- \kappa + \gamma_{1} 
\end{array} \right) 
\label{jacobian} 
\end{equation}

The calculation of the convergence and shear can serve
as an important tool to distinguish between braneworld gravity and GR. In order 
to calculate these quantities
for a non-zero $\Psi$, we use the 
 spherical symmetry ($\theta_1 = \theta_2 = \theta$) of the lens, which yields
\begin{eqnarray}
\kappa &=& \frac{1}{2} \text{Tr} (V_{R ij} - V_{\Psi ij}) 
= \frac{\partial^2 (V_R - V_\Psi)}{\partial \theta^2} \label{kappa} \\
\gamma_1 &=& \frac{1}{2}\left[(V_{R 11} - V_{\Psi 11})- (V_{R 22} - V_{\Psi 22})\right]
= 0 \label{gamma1} \\
\gamma_2 &=& V_{R 12} - V_{\Psi 12} = V_{R 21} - V_{\Psi 21}
= \frac{\partial^2 (V_R - V_\Psi)}{\partial \theta^2} \label{gamma2} \\
\gamma &=& \gamma_2 =  \frac{\partial^2 (V_R - V_\Psi)}{\partial \theta^2} \label{gamma} 
\end{eqnarray}
The results show that both the convergence and
the shear are less than the corresponding GR values due to the presence of a 
non-zero relativistic potential.

We can now construct the Jacobian matrix by using its components 
as calculated above. Separating the braneworld modifications from the GR values, we 
finally arrive at
\begin{equation}
 {\cal A} =
\left( \begin{array} {cc}
  1 - \kappa_R - \gamma_{1R}& -\gamma_{2R} \\
  - \gamma_{2R}&  1- \kappa_R + \gamma_{1R} 
\end{array} \right) +
\left( \begin{array} {cc}
    \kappa_\Psi + \gamma_{1 \Psi}& \gamma_{2 \Psi} \\
   \gamma_{2 \Psi}&   \kappa_\Psi - \gamma_{1 \Psi} 
\end{array} \right)
\label{brjacobian}  
\end{equation}
The above equation  shows explicitly the role the relativistic potential plays in 
determining the magnification. The first matrix is the Jacobian in GR  while the
second one is the exclusive contribution from a non-zero relativistic potential. 
This expression clearly reveals that the determinant of the Jacobian with a non-zero $\Psi$ is 
different from the GR value (where $\Psi =0$).
However, whether this determinant will have a positive or a negative contribution solely
depends upon the explicit expression for the relativistic potential one gets on solving
the effective Einstein equation. Thus, though the convergence and shear are
less than the GR value due to a positive $\Psi$, the expression for  
the magnification being highly nonlinear, one cannot say {\em a priori} 
whether the magnification is more or less than GR. What one can say at best
is that the magnification will be different from GR. It is only when one has
a specific expression for $\Psi$, one can  
calculate this difference ({\em i.e.}, more or less) conclusively,
 a fact which resonates with the discussions following Eq (\ref{mag2}).
In the following section, we shall calculate these quantities for specific potentials
and estimate the difference of the quantities  from GR.

Thus, we arrive at the conclusion that \textit{finding out the magnification
by  spherically symmetric lenses  by measuring the convergence and shear 
can help us test braneworld gravity, and in general, any theory of gravity
with two potentials, through observations.}


\section{Quantitative estimation}

Let us now try to make some actual quantitative estimation of  
lensing effects by clusters and spiral galaxies on the brane and see by how much
amount the observable quantities differ from the GR values.
To this end, we shall make use of the Newtonian 
and relativistic potentials obtained in \cite{clust, altdm}.


\subsection{Lensing by clusters}

\noindent For an X-ray cluster on the brane, we employ the Newtonian and relativistic potentials 
obtained in \cite{clust}. Upon scaling
with $c$, they read
\begin{eqnarray}
\Phi(r) &=& \frac{2 k T}{\mu m_p} \ln\frac{r}{r_c} \label{phiclust} \\
\Psi(r) &=& \left[\frac{k T}{\mu m_p} -2 \pi G \rho_0 r_c^2\right]  \ln\frac{r}{r_c}
\label{psiclust} 
\end{eqnarray}
where $\rho_0, ~ r_c, ~\mu, ~T$ are respectively the central density,
core radius, mean atomic weight of gas particles inside the X-ray cluster and the
temperature of the gas.

In the standard GR analysis of X-ray profiles of clusters by using dark matter,
 $\Psi=0$ and the deflection angle $\hat{\alpha}_R$ of a photon 
from a distant  source, propagating through the dark matter halo  to a distant
observer  is given by
\begin{equation}
\hat{\alpha}_R = \frac{2}{c^2} \int_{S}^{O}  \hat{\nabla}_\perp  \Phi
\, \,  dl \, 
\label{grdef} 
\end{equation}
Using the above expression for $\Phi$, we find from GR that a photon passing
through the halo of a cluster experiences a constant deflection  
\begin{equation}
\hat \alpha_R=\frac{4\pi kT}{\mu m_p c^2}
\end{equation}
In braneworld gravity $\Psi \neq 0$ and  the deflection angle is now modified to
Eq (\ref{def}). For a cluster with  the above
 $\Phi$ and $\Psi$ as calculated from braneworld gravity,  this deflection angle turns out 
to be 
\begin{equation}
\hat{\alpha } = \frac{3\pi kT}{\mu m_p c^2} + \frac{2 \pi^2 G \rho_0 r_c^2}{c^2}
= \hat{\alpha_R} \left[\frac{3}{4} + \frac{\pi G \rho_0
    r_c^2 \mu m_p}{2 k T}\right]  
\end{equation}
For a typical X-ray cluster, we use the
following representative values for the cluster parameters \cite{clusterd}: 
$\rho_0 = 5 \times 10^{-24} {\rm kg/m^{3}}, ~
 r_c = 0.3 {\rm Mpc}, ~
 \mu = 0.6, ~
T = 10^{8} {\rm K}$.
A good summary of up-to-date cluster data is also available in \cite {clusterdsum}
for further confirmation of these data.
Consequently, the deflection angle from braneworld gravity turns out to be
 around $\sim 80 \%$ of the GR value.

As already pointed out, the different observable properties of  
lensing for a cluster in the brane
will also differ significantly from the GR values. 
Below we mention the estimates for some of the
observable quantities, namely convergence and shear,
 for an X-ray cluster with our choice of parameters.

\begin{table} [htb]
\begin{center} 
\begin{tabular}{|c|c|c|}

\hline \textbf{Properties} & \textbf{Estimations} & \textbf{Comments} \\

\hline Image position &$\theta = \theta_{R} \left[\frac{3}{4} + \frac{\pi G \rho_0
    r_c^2 \mu m_p}{2 k T}\right]$ &  single image, closer by $20 \%$\\

\hline Convergence & $\kappa = \kappa_R \left[\frac{3}{4} + \frac{\pi G \rho_0
    r_c^2 \mu m_p}{2 k T}\right]$  &   $20 \%$ less change in image size\\

\hline Shear & $\gamma_1 = 0 = \gamma_{1R}$ & $\gamma_1$ unchanged\\ 
& $\gamma_2 = \gamma_{2R}\left[\frac{3}{4} + \frac{\pi G \rho_0
    r_c^2 \mu m_p}{2 k T}\right] $ & $ \gamma_2$ less by $20 \%$\\
&$\gamma = \gamma_{2}$ & $\Rightarrow$ change in shape $20 \%$ less\\

\hline

\end{tabular}
\end{center}
\caption{A comparative analysis of different observable properties of gravitational 
lensing by a cluster obtained from  braneworld gravity with their GR counterparts
for $\beta =0$.}
\end{table}
We find that there is a $\sim 20 \%$ difference
 in the estimation of  these observable quantities in  lensing  
 in the two different theories. The results can be 
compared with observations in order to test braneworld gravity using the formalism.


\subsection{Lensing by spiral galaxies}

\noindent As another interesting situation where we can test braneworld gravity, we intend
to estimate the  lensing effects  for a spiral galaxy on the brane.
For explicit calculations,
we take up the Newtonian and relativistic potentials 
 found in \cite{altdm} by scaling with $c$ 
\begin{eqnarray}
\Phi(r) &=& v_c^2 \left[\ln \left(\frac {r}{r_0} \right)-1 \right] \label{phigalaxy} \\
\Psi(r) &=& \frac{v_c^2}{2} \left[\ln \left(\frac {r}{r_0} \right)-1 \right]
- \left[\frac{4 \pi^2 G \rho_0}{\gamma^2}\right] {1 \over r} \label{psigalaxy} 
 \end{eqnarray}
where $v_c, ~ r_0, ~\rho_0$ are respectively the rotational velocity in the flat rotation curve region,
the impact parameter and the core density. 
 
In the GR analysis of rotation curves of spiral galaxies, 
the GR deflection angle  of a photon
is determined by  Eq (\ref{grdef}).
Consequently, the deflection angle of a photon passing through
the galactic halo turns out to be 
\begin{equation}
\hat{\alpha}_R = \frac{2 \pi v_c^2}{c^2}  
\end{equation}
which is nothing but the deflection angle for a singular isothermal sphere in GR, 
whereas for the galactic metric obtained from braneworld gravity for a non-zero $\Psi$,
 the deflection angle is found to be
\begin{equation}
\hat{\alpha } = \frac{3\pi v_c^2}{2 c^2} - \frac{8 \pi^2 G \rho_0}{\gamma^2 c^2 b}
=\hat{\alpha_R} \left[ \frac{3}{4} 
- \frac{4 \pi G \rho_0}{\gamma^2 v_c^2 b}\right]  
\end{equation}
where $b$ is the usual impact parameter.
For estimation, we use the following values of the parameters
for a typical spiral galaxy \cite{binney} : 
$v_c = 220 {\rm km/s}, ~
r_0 = 8 {\rm kpc} ~(\sim \gamma^{-1} \sim b), ~
\rho_0 = 10^{-25} {\rm kg/m^2}$ (note that $\rho_0$ is the surface density).
Thus, the deflection angle by a galaxy in the braneworlds turns out to be $\sim 75 \%$ of
the GR value. 

Likewise, the other observable properties for gravitational lensing by a galaxy
can also be estimated and
compared with their GR counterparts
by noting the fact that the impact parameter is related to
the angular position of the image by $b \propto \theta$.
 The following table summarizes the results.
\begin{table} [htb]
\begin{center} 
\begin{tabular}{|c|c|c|}

\hline \textbf{Properties} & \textbf{Estimations} & \textbf{Comments} \\ 

\hline Image position &$\theta_{+} =
\theta_{R} \left[\frac{3}{4} 
- \frac{8}{3}\frac{G \rho_0 D_s c^2}{D_d D_{ds}\gamma^2 v_c^4} \right]  $ &
 image closer by $25 \%$ \\
& $\theta_{-} = \frac{8}{3}\frac{G \rho_0 D_s c^2}{D_d D_{ds}\gamma^2 v_c^4}$
& second image closer to $\theta=0$ \\ 

\hline Convergence & $\kappa = \kappa_R \left[\frac{3}{4} + \frac{2 \pi G \rho_0}{\gamma^2 v_c^2}
\frac{\xi z}{(\xi^2 + z^2)^\frac{3}{2}}\right ]{\vert}_{-D_{ds}}^{D_d}$  &   $25 \%$ change in image size\\

\hline Shear & $\gamma_1 = 0 = \gamma_{1R}$ & $\gamma_1$ unchanged\\ 
& $\gamma_2 = \gamma_{2R} \left[\frac{3}{4} + \frac{2 \pi G \rho_0}{\gamma^2 v_c^2}
\frac{\xi z}{(\xi^2 + z^2)^\frac{3}{2}}\right] {\vert}_{-D_{ds}}^{D_d} $ & $ \gamma_2$ less by $25 \%$\\
&$\gamma = \gamma_{2}$ & $\Rightarrow$ change in shape $25 \%$ less\\

\hline

\end{tabular}
\end{center}
\caption{A comparative analysis of different observable properties of weak 
lensing by a spiral galaxy in  braneworlds with their GR counterparts for $\beta =0$.
Here $\xi$ and $z$ are, respectively, the projected radius along the impact parameter and 
the path length of the light ray.}
\end{table}

In a nutshell,  the quantities differ by $\sim 25 \%$ from GR, which is good enough
to distinguish between the two theories. 
The result can again be subject to observational verification to test braneworld gravity
theory.



\subsection{Present status of observations}

We have shown that sufficiently accurate  lensing data for
clusters and galaxies can be useful to test braneworld gravity. 
The present observational  data \cite{clusterlens1, clusterlens2}
reveal that there are significant amount of uncertainties 
in the galaxy or cluster properties estimated from the lensing data.
While a few of them claim that they are consistent \cite{clusterlens2},
some of them \cite{clusterlens1} indeed show that there are some inconsistency between 
the observation and the theory based on dark matter.
 The uncertainty in these data thus opens up a fair
possibility for a modified theory of gravity, \textit{e.g.},
braneworld gravity, to replace GR in explaining those observations. 
For example, lensing calculations from the nonsymmetric theory of gravity 
\cite{moffat} has also shown its possibility to be an alternative to GR in
galactic and extragalactic scales.

Using weak lensing data, the best fit velocity dispersion for a cluster
has been found to be $2200\pm500 {\rm km/s}$. 
Analyzing the change in the background galaxy luminosity function, the 
cluster mass is obtained in the range $(0.48\pm 0.16)\times 
10^{15} h^{-1} M_{\odot}$ at a radius $0.25 h^{-1}$ from the cluster core \cite{datamass1}. 
Further information about the determination of mass can be obtained from \cite{datamass2, datamod}.
Magnification \cite{datamag} and shear \cite{datashear} can also be calculated
from the data. For example,  \cite{datashear} estimates the amount of shear for
a typical cluster to be $\langle \gamma^2\rangle^{1/2} = 0.0012 \pm 0.0003$.
These results reveal $\sim 25 - 30 \%$ uncertainties in determining the precise value
of the quantities.

Several properties of galaxy dark matter halos can be
derived from weak lensing \cite{lensdata1, lensdata2}. 
Using the galaxy-mass cross-correlation function, it is found that
the value of velocity dispersion is $\langle\sigma_v^2 \rangle^{1/2} =  128 \pm 4
{\rm km/s}$ \cite{lensdata1}. But this value is highly sensitive to the selection of
the sample of lens galaxies, \textit{e.g.}, with different samples, the value
lies in between $118\pm4\pm2 {\rm km/s}$ and $140\pm4\pm3 {\rm km/s}$. Thus the 
results are not so precise. 
A detailed survey of the current status of weak lensing can by found in \cite{datarev}.
 
To conclude, at the present status of informations,
  both GR and braneworld gravity would fare equally well 
in explaining those observations.
The results showing the present status of weak lensing are thus insufficient for a conclusive remark. 
 A more accurate measurement of those lensing effects  will
help us determine conclusively  whether or not braneworld gravity can be accepted as the theory of
gravity.


\section{Summary and outlook}

We have developed a formalism appropriate for 
understanding gravitational lensing in the line elements which arise in 
braneworld gravity. Of course, this formalism is general enough for
studying lensing in contexts wherever two gravitational
potentials are required in order to include relativistic effects. 
For instance, following earlier work, one may use
our general formulae for studying dark matter scenarios where 
pressure is not negligible {\cite{sbsk}}. 
 
With the intention of studying gravitational lensing in detail, we have 
obtained, using our formalism,  general expressions for the time delay, 
deflection angle, Einstein ring, image positions, magnification and critical 
curves.
It was noted that significant deviations from the results of weak--field 
GR was evident in the expressions for each of the abovementioned quantities. 

To illustrate our formalism, we  made use of our earlier results
on gravitational potentials of clusters and spiral galaxies, 
as obtained in braneworld gravity (using 
the relativistic, but weak--field effective Einstein equations on the brane). 
We estimated quantitatively lensing features 
for clusters and spiral galaxies
by using both the Newtonian and weakly relativistic potentials. 
The difference between
the values of each of the above quantities as compared to those
obtained in the standard scenario, is found to be around $20 - 25 \%$. 
Analysis of 
actual data reveals a $25 - 30 \%$ uncertainty in the values of almost all
of these quantities. Thus, we conclude 
that it is only
when  more precise data become available, the theory can be verified 
conclusively, using lensing observations.

In this article, we have primarily focused on weak lensing effects 
which can act as signatures for a modified theory of gravity. 
It is surely worthwhile to investigate features of strong
lensing as well, which may provide further 
ways of testing braneworld gravity, or,
for that matter, any modified theory of gravity where a two potential
formalism becomes necessary. To this end,
we have performed some simplistic calculations of caustics and 
critical curves, assuming a spherically symmetric lens 
considered as a singular isothermal sphere,
and have obtained some preliminary results.
The critical curves have been found to give qualitatively same but 
quantitatively different, 
though the location of the caustics remain unchanged.
Thus, we expect, that a detailed survey of strong lensing
in braneworld gravity may reveal further interesting and new features. 
We hope to address such issues related to strong lensing 
in detail, in future. 

In conclusion, it is important to mention a drawback in our
formalism. The general results we have obtained are applicable only to 
lensing by local objects in the sky. We need to include the effects of a 
background cosmology in order to address more realistic scenarios in an 
appropriate manner. We hope to return to this and other issues later.

\section*{Acknowledgments}

We thank S. Bharadwaj for discussion and suggestions related to the work 
reported in this article. We also acknowledge  useful discussions with 
S. Majumdar, R. Misra, T. Padmanabhan, T. D. Saini and K. Subramanian.
Thanks also to Dibyendu Mandal for pointing out a correction in the estimation.


\end{document}